# A Kapitza-Dirac-Talbot-Lau interferometer for highly polarizable molecules

Stefan Gerlich[1], Lucia Hackermüller[1*], Klaus Hornberger[2], Alexander Stibor[1†], Hendrik Ulbricht[1], Michael Gring[1], Fabienne Goldfarb[1‡], Tim Savas[3], Marcel Müri[4], Marcel Mayor[4,5] and Markus Arndt[1$]

1 Faculty of Physics, University of Vienna, Boltzmanngasse 5, A-1090 Wien
2 Arnold Sommerfeld Center for Theoretical Physics, Ludwig-Maximilians-Universität München, Theresienstraße 37, 80333 München, Germany
3 Nanostructures Laboratory, Massachusetts Institute of Technology, Cambridge, MA
4 University of Basel, Department of Chemistry, St. Johannsring 19, CH-4056 Basel, Switzerland
5 Forschungszentrum Karlsruhe GmbH, Institute for Nanotechnology, P. O. Box 3640, D- 76021 Karlsruhe, Germany
* Present address: Johannes Gutenberg-Universität Mainz, Staudingerweg. 7, 55099 Mainz
† Present address: Phys. Institut der Universität Tübingen, Auf der Morgenstelle 14, D-72076 Tübingen
‡ Present address: Laboratoire Aimé Cotton, CNRS-UPR 3321, Bat. 505, Campus Univ., F-91405 Orsay
$ e-mail: markus.arndt@univie.ac.at

**Research on matter waves is a thriving field of quantum physics and has recently stimulated many investigations with electrons [1], neutrons [2], atoms [3], Bose-condensed ensembles [4], cold clusters [5] and hot molecules [6]. Coherence experiments with complex objects are of interest for exploring the transition to classical physics [7-9], for measuring molecular properties [10] and they have even been proposed for testing new models of space-time [11].**
**For matter-wave experiments with complex molecules, the strongly dispersive effect of the interaction between the diffracted molecule and the grating wall is a major challenge because it imposes enormous constraints on the velocity selection of the molecular beam [12]. We here describe the first experimental realization of a new interferometer that solves this problem by combining the advantages of a Talbot-Lau setup [13] with the benefits of an optical phase grating and we show quantum interference with new large molecules.**

Several methods have been developed in the past for the coherent manipulation of matter-waves with de Broglie wavelengths in the nanometer and picometer range. Free-standing material gratings were for instance used in the diffraction of electrons [14], atoms [15,16] and molecules [5,6,17]. Also the coherent beam splitting at non-resonant standing light waves, often designated as the Kapitza-Dirac effect, has been observed for all of these species [18,19,20].

Recent implementations of near-field interferometry [13,21,22,23] underlined the particular advantages of the Talbot-Lau concept for experiments with massive objects: The required grating period scales only weakly ($d \sim \sqrt{\lambda}$) with the de Broglie wavelength, and the design accepts beams of low spatial coherence which makes high signals possible even for weak sources.
A symmetric Talbot-Lau interferometer (TLI) consists of three identical gratings. The first one prepares the transverse coherence of the weakly collimated beam. Quantum near-field diffraction at the second nanostructure generates a periodic molecular density distribution at the position of the third mask, which represents a self-image of the second grating, if the grating separation equals a multiple of the Talbot length $L_T = d^2/\lambda$. The mask can be laterally shifted to transform the molecular interference pattern into a modulation of the molecular beam intensity that is recorded behind the interferometer.

In the established TLI design with three nanofabricated gratings [23], the molecule-wall interaction with the grating bars imprints an additional phase shift φ on the matter wave, which depends on the molecular polarizability α, the velocity $v_z$ and the distance $r$ to the wall within the grating slit. Because of its strongly non-linear $r$-dependence this interaction restricts the interference contrast to very narrow bands of de Broglie wavelengths, as we illustrate in Fig. 1a for the example of the fullerene $C_{70}$. In this simulation we utilize the full Casimir-Polder potential [24], even though the long-distance (retarded) approximation, decaying as $α/r^4$, closely reproduces the results. The sharply peaked thin line shows the expected interference fringe visibility as a function of the de Broglie wavelength for a Talbot Lau interferometer composed of three silicon nitride gratings with a period of 266.38 nm. The periodic recurrence of the fringe visibility with multiples of $\lambda=d^2/L$ is a generic feature of any such three-grating setup. The grating separation was fixed to L=105 mm. The left peak in Fig. 1a therefore corresponds to the 4$^{th}$ Talbot order for $C_{70}$ at 175 m/s. The presence of the walls modifies the transmission function [25] and leads to a dramatic narrowing of the accepted width of the wavelength distribution, here corresponding to a velocity spread of $\Delta v/v_z$ =0.7% (FWHM). This requirement has to be contrasted with available molecular beam methods: currently available sources for very massive neutral molecules only allow to prepare beams which exhibit either a velocity or mass spread (or both) of often significantly more than 10%. A better post-selection is conceivable in principle, but the existing sources provide generally an insufficient flux to allow this.

For more complex particles with larger polarizabilities, lower velocity and higher velocity spread than $C_{70}$, the phase shift in the presence of material walls will even be more important. We demonstrate this in Fig. 1b, where we simulate the expected fringe visibility for large perfluorinated molecules [26] in a conventional TL-interferometer. The presence of dispersive interactions clearly reduces the expected fringe contrast to values around 4% (thin line, Fig. 1.b1).

In order to circumvent the molecule-wall interaction, we now combine the Talbot-Lau concept with the idea of the Kapitza-Dirac effect, i.e. with a standing light wave as the diffraction element.

Specifically, we replace the central grating by a standing laser light wave of period $d=\lambda_L/2$, which interacts with the molecules by the dipole force [20]. The spatially varying laser intensity thus induces an oscillating dipole moment shifting the phase of the matter wave in proportion to the local intensity and to $α_L$, the molecular polarizability at the laser wavelength $\lambda_L$. A phase grating constructed this way has several advantages: it is indestructible and it combines 100% transmission with a tunable phase shift. But most importantly, all parameters can be chosen in order to obtain a high interference contrast over a broad wavelength distribution as simulated in Fig. 1.a2 (dotted line).

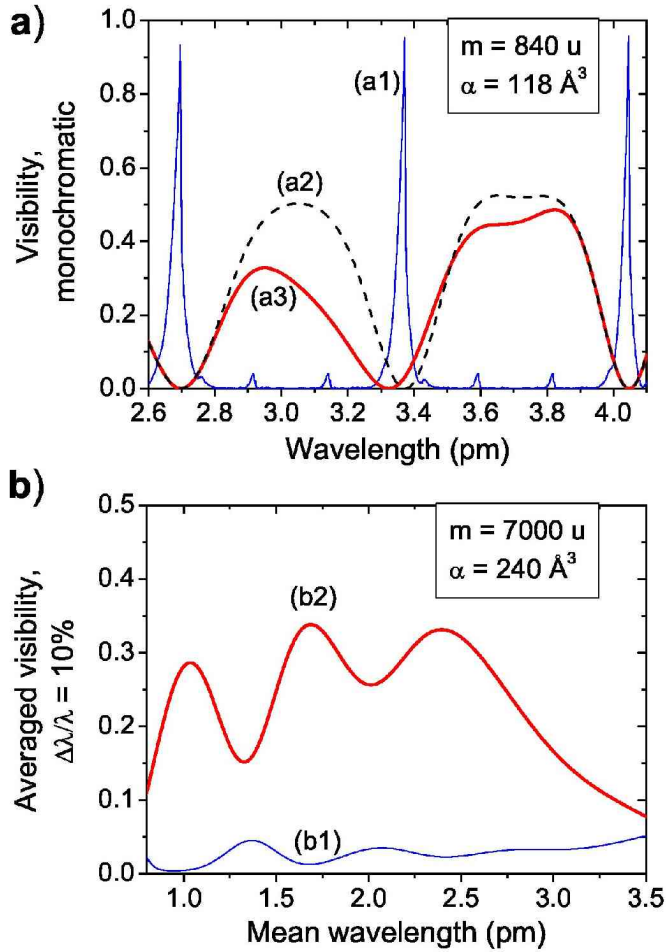

*Figure 1: Numerical prediction of the interference fringe visibility as a function of the de Broglie wavelength: Comparison of the Talbot-Lau – interferometer (TLI) and the Kapitza-Dirac-Talbot-Lau-interferometer (KDTLI) for molecules of different complexity:*

- **(a)** *Fullerene $C_{70}$ contrast* under the assumption of a monochromatic beam. The thin solid line (a1) shows the expected fringe visibility for a pure TLI setup, i.e. when only silicon nitride gratings are used in all three places. Only in a very narrow band of de Broglie wavelengths ($\Delta\lambda/\lambda < 0.7\%$ FWHM) the visibility can reach significant values. In marked contrast to that, the $\Delta\lambda$ acceptance range of the KDTLI is more than ten times larger (dashed curve, a2), even in the presence of a realistic photon absorption rate (solid curve, a3). These curves are based on Eq. (3) with P=5 W and $\sigma_{abs}=2.1 \times 10^{-21}$ m$^2$. A wavelength average over a realistic velocity distribution of 10% (standard deviation) leads to a total fringe contrast of <5% for the TLI and up to 25% for the KDTLI setup.
- **(b)** *Perfluorinated macromolecules:* The interference visibility in this panel is wavelength averaged over an experimentally realistic distribution of mean velocities in the range of $v_z$=20-70 m/s with $\Delta v/v_z = 10\%$. For such slow and polarizable particles, the fringe visibility of the KDTLI design (b2) at P = 0.6 W and $\sigma_{abs} = 1.34 \times 10^{-21}$ m$^2$, can clearly exceed that of the TLI design (b1) by about a factor of ten.

We note that the first and third grating may remain absorptive masks. The first grating selects a periodic set of slit sources from the molecular beam and thus prepares the required transverse coherence [23]. Similarly, the third mask provides the high spatial resolution in the detection of the periodic molecular interference pattern.

The molecule-wall interaction is also of minor relevance in both the entrance and the exit grating: the molecules enter the first grating without a well-defined phase and we observe only molecular flux, and not phase, behind the third nanostructure.

The molecule-laser interaction can be characterized by the phase $\Phi_{max}$ that is imprinted on the de Broglie wave at the position of the maxima of the standing light wave [20]

$$\Phi_{max} = \frac{8\sqrt{2\pi}\,\alpha_L}{\hbar c w_y v_z} P \quad . \tag{1}$$

This amounts to $\Phi_{max} \sim \pi$ for $C_{70}$ with a polarizability of $\alpha_L = 118$ Å$^3$ at $\lambda_L = 532$ nm, a molecular velocity of $v_z = 130$ m/s, a vertical laser beam waist of $w_y = 900$ µm and a laser power of $P = 5$ W.

In addition to this coherent effect of the laser potential we also have to consider the possibility of photon absorption. Each absorbed photon adds a momentum kick of $\pm h/\lambda_L$ to the molecules and thus adds to the blur of the accumulated interference pattern. This incoherent interaction is characterized by the mean number of absorbed light quanta in the antinode

$$n_0 = \frac{8\sigma_{abs}\lambda_L}{\sqrt{2\pi}\,hc w_y v_z} P \tag{2}$$

and amounts in our example to $n_0 = 0.8$ photons per molecule, with the absorption cross section $\sigma_{abs} = 2.1 \times 10^{-21}$ m$^2$ [28] and all other parameters as above.

A rigorous treatment within the framework of our previously established phase-space theory for the Talbot-Lau interferometer [23,27] predicts that the interference pattern behind the KDTLI will be described by a nearly sinusoidal curve of visibility

$$V = 2\left(\frac{\sin(\pi f)}{\pi f}\right)^2 \exp(-\xi_{abs})\frac{\xi_{coh}-\xi_{abs}}{\xi_{coh}+\xi_{abs}} J_2\left(-sgn(\xi_{coh}+\xi_{abs})\sqrt{\xi_{coh}^2-\xi_{abs}^2}\right) \quad . \tag{3}$$

Here the coherent diffraction parameter

$$\xi_{coh} = \Phi_{max} \sin\left(\pi\frac{L}{L_T}\right) \tag{4}$$

and the parameter of absorption

$$\xi_{abs} = n_0 \sin^2\left(\frac{\pi L}{2 L_T}\right) \tag{5}$$

are determined by the grating separation L, the Talbot length $L_T$ and the open fraction $f = 0.42$, i.e. the ratio of slit size and period in the nanofabricated gratings. $J_2$ designates the Bessel function of second order. In the limit of vanishing absorption ($\xi_{abs} \to 0$) it reduces to $J_2(\xi_{coh})$.

The thick line of Fig. 1a shows the full effect of the laser interaction according to Eq. (3). One observes that the interference visibility is reduced, in particular for slow molecules which have more time to absorb a photon. Yet, the most important advantage of the KDTLI design, a significant interference contrast over a broad wavelength range, is clearly preserved.

We note that for many large molecules, from perfluorinated particles to large polypeptides, the absorption lines are blue shifted with respect to those of fullerenes, and the ratio $\sigma_{abs}/\alpha_L$ at 532 nm can be more than ten times smaller than for $C_{70}$ [28]. For these objects the effect of absorption will be negligible and even higher fringe visibilities can be expected. This is again demonstrated in Fig. 1b, where we simulate the interference contrast in our KDTLI setup for perfluorinated molecules with m=7,000 amu. A comparison between this result (solid line, Fig 1.b2) and the prediction for purely matter gratings (thin line, Fig. 1.b1) reveals that our new setup allows to gain one order of magnitude in fringe visibility.

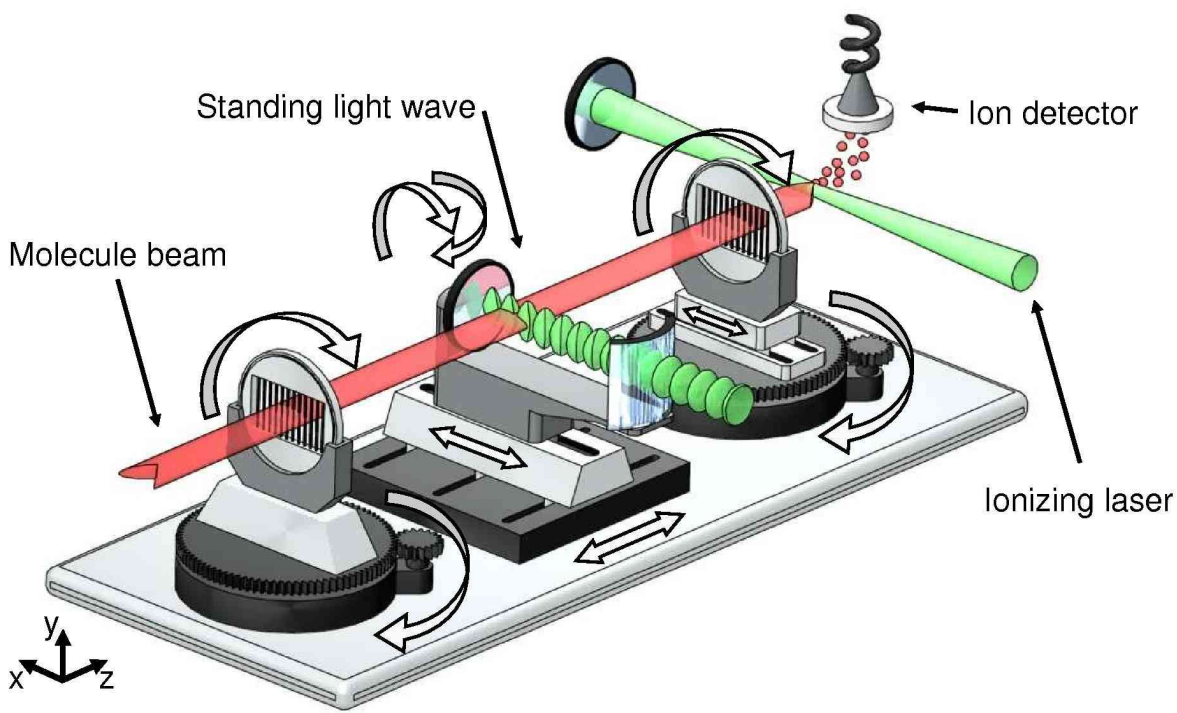

*Figure 2: Artist's view of the Kapitza-Dirac-Talbot-Lau-Interferometer*
The molecular beam, arriving from the left, passes the first nanograting which imposes a periodic spatial filtering and thus the required transverse coherence onto the molecular beam. The spatially varying potential of the standing light leads to a phase modulation of the matter wave, which results in a molecular density pattern at the position of the third grating. The latter is again a material mask of equal period and can be laterally shifted across the molecule beam to sample the interference pattern. High-vacuum motors are used to fine-adjust the roll angle of all gratings, their longitudinal separation, as well as the transverse shift of the third grating.
$C_{70}$-Fullerenes are detected by laser ionisation, as depicted here. Perfluoroalkyl-functionalized azobenzenes are recorded using electron impact ionisation mass spectrometry.

In order to validate the concept we have implemented the interferometer for $C_{70}$ as depicted in the artist's view of Fig. 2. The molecular beam emerges from a thermal source. Its mean velocity can be selected using a previously established gravitational selection scheme [23] within the range of $v_z$=80...200 m/s and with $\Delta v_z/v_z$ ~ 7...20 % (standard deviation). The molecules which pass the third grating are ionized by a green 18 W laser beam, which is focused to a waist of 15 µm in a double pass arrangement. They are subsequently detected in an ion counter.

The free-standing gratings were photo-lithographically etched into a 190 nm thick silicon nitride ($SiN_x$) membrane [29]. A crucial point in the preparation of the experiment is the precise matching of all three grating periods. Already an average deviation by as little as 0.05 nm, i.e. by the radius of a single hydrogen atom (!), reduces the fringe visibility by one third.

In the experiment we sample the interference curve by shifting the third grating laterally while counting the total number of transmitted molecules. Figure 3a shows exemplary interference fringes recorded with $C_{70}$ in the KDTLI setup.

From several such curves we then extract the fringe visibility of the interference signal as a function of the grating laser power. In Fig. 4 we compare the result with the predictions of equation (3), including the measured velocity spreads, for four different mean molecular velocities. The experimental data validate our theoretical model to a very high degree and thereby also confirm the enormous precision in the manufacturing of the diffraction elements. Depending on the laser power we reach an interference contrast of up to 24%. This represents a significant improvement over a Talbot-Lau interferometer with three material gratings, for which the visibility would remain below 6% at these broad velocity distributions.

The new instrument now allows us to demonstrate quantum interference with a new class of molecules, here with perfluoroalkyl-functionalized azobenzenes, which were synthesized for this specific purpose (see methods and supplementary information). In their thermally preferred *trans*-conformation [30] (inset of Fig. 3b) they are about four times longer and also a bit more massive (~1030 amu) than the fullerene $C_{70}$. Although the limited count rate forced us to work with a rather broad, nearly thermal velocity distribution ($\Delta v/v_z \approx 0.25$), we still see a good interference contrast which is very well compatible with our quantum interference model (Fig. 3b).

All parameters of our present experimental setup are still compatible with molecular beams in a mass range up to 11,000 amu and velocities of 50 m/s. Such experiments should become accessible in the near future by combining modern chemical synthesis with improved beam sources. When we extrapolate the simulations for our setup even to $Au_{5000}$ clusters at cryogenic temperatures (~10 K) we still predict a fringe visibility of 40% for a mass of m~1,000,000 amu, a polarizability of α~25,000 Å$^3$, velocities in the range of $v_z$ ~1 m/s and a mass or velocity spread of up to 10%. Such an experiment represents a very worthy goal for future research, with interesting challenges related to the enormous sensitivity of such an interferometer to alignment, inertial forces and dephasing effects.

The Kapitza-Dirac-Talbot-Lau concept thus opens de Broglie interferometry to a very wide class of clusters and molecules in an unprecedented mass and complexity region.

Our demonstration with perfluoroalkyl-functionalized azobenzenes underlines in particular that such experiments can be performed with realistic molecular beams, i.e. also with a rather broad velocity distribution.

The new interferometer also leads to a number of applications, including tests of recent proposals on gravitational decoherence [11] or molecule metrology [10]. It is particularly interesting to note that azobenzenes and their derivatives are often used as molecular switches, which change their conformation upon absorption of a single photon [30]. It will therefore be intriguing to study the possibility of optically controlled conformational state changes on quantum interferometry and decoherence in the future.

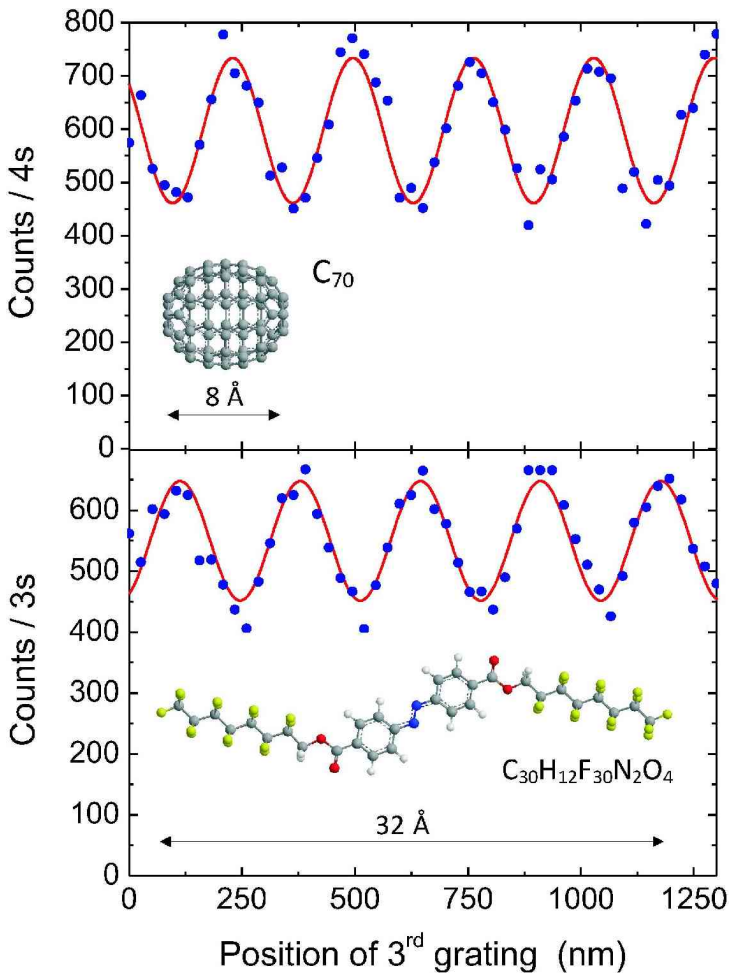

*Figure 3: Typical interference patterns observed in KDTL interferometry*

(a) Interference fringes of $C_{70}$, recorded at a mean molecular velocity of 146 m/s ($\Delta v/v_z$ = 16 %) and a diffraction laser power of 6 Watts. The solid line is a sinusoidal fit. The observed fringe visibility of 23 % is in good agreement with the expected values of Fig. 1 and Fig. 4, when the appropriate velocity averaging is taken into account.

(b) Quantum interference with perfluoroalkyl-functionalized azobenzenes. The laser power was set to 7.3 Watts. The velocity distribution was centred at $v_z$ = 140 m/s with a standard deviation of 25%. A dark count rate of 15 counts per second was subtracted. The experimental fringe visibility of 18.0 % is in good agreement with the quantum theoretical expectation (18.5%) and three times higher than expected for a purely mechanical Talbot Lau interferometer. The fringe period is determined by the geometry of the setup and equal for both the fullerenes and the azobenzenes.

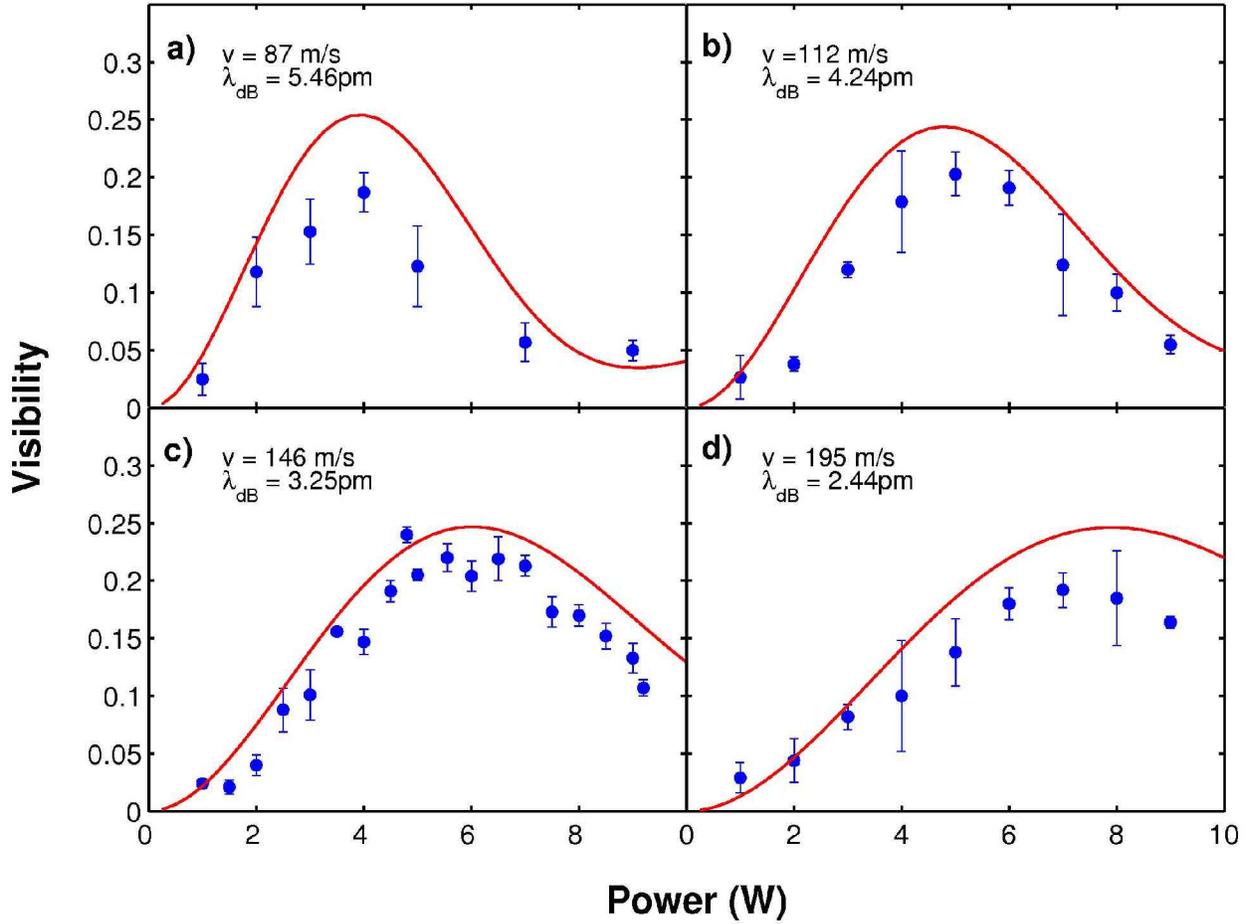

*Figure 4: Dependence of the interference fringe contrast on the diffracting laser power*
We show the $C_{70}$ interference visibility as a function of the diffraction laser power at four different molecule velocity distributions, with mean velocities: a) 87 m/s, $\Delta v/v_z = 0.09$, b) 112 m/s, $\Delta v/v_z = 0.13$, c) 146 m/s, $\Delta v/v_z = 0.16$, d) 195 m/s, $\Delta v/v_z = 0.21$ (standard deviation).
The circles mark the experimental results. The error bars correspond to the standard deviation of three consecutive measurements. The solid lines represent the theoretical expectation of equation (3), including both the dispersive and the absorptive effects on the molecule-light interaction without any free parameters. We attribute the small remaining deviations between experiment and theory mainly to day-to-day variations of the vibrational noise in the lab.
The centre of the peak shifts towards higher laser power for increasing velocities. This is related to the decreasing interaction time with the laser field.

**Methods:**

*Experimental layout and alignment of the interferometer:*
The KDTLI interferometer is composed of three gratings, two absorptive silicon nitride gratings and one optical phase grating. The optical grating was created by focusing a green, single-mode laser beam (Coherent Verdi 10 W) with a cylindrical lens to a horizontal waist of $w_z=20$ µm onto a plane mirror, to create a standing light wave. The Rayleigh length of the Gaussian laser beam amounts to 2.4 mm, which is sufficiently long to keep the laser wave front curvature small over the intersection region with the molecular beam. The vertical waist was set to $w_y=900$ µm to maximize the field homogeneity over the interaction region. The laser has a fixed wavelength of 532.28 ± 0.01 nm, and the material gratings had to be tailored to half this value. They were manufactured to d=266.38 ± 0.05 nm and the effective grating period was tuned in situ by rotating the gratings around their vertical axis (see Fig. 2).

The molecular beam divergence and the beam alignment with respect to the mirror plane are restricted to about 1 mrad to avoid phase averaging over different nodal planes of the light wave. All gratings were optically aligned to be parallel to each other to better than 300 μrad. The separations between two neighbouring gratings differ by less than 50 μm. Their distance was set to L=105 mm. This corresponds to the first Talbot length for molecules with m=11,000 amu and $v_z$ =50 m/s ($\lambda_{dB}$ =700 fm). For $C_{70}$ it amounts to the 4$^{th}$ up to the 8$^{th}$ Talbot order, depending on the selected velocity. To correctly include the average over a finite velocity distribution the signal average is computed before we extract the corresponding visibility.

*Synthesis, properties and detection of the azobenzene derivative functionalised with fluorinated alkyl chains.*
To minimize intermolecular attraction and hence to increase the compound's vapour pressure, an azobenzene derivative with fluorinated alkyl chains was synthesized by esterification of azobenzene 4,4'-dicarboxylic acid with a highly fluorinated alkyl alcohol (synthetic protocol in supplementary information). Azobenzene 4,4'-dicarboxylic acid was obtained in 83% by treating 4-nitrobenzoic acid with sodium hydroxide, D-glucose and air in water. The corresponding acid chloride derivative was obtained by treatment with thionyl chloride. Subsequent reaction with 1H-1H-perfluoro-1-octanol in dry THF and triethylamine as base afforded the desired azobenzene 4,4'-di(carboxylic acid 1"H-1"H-perfluoro-1"-octanolate) as orange crystalline solid. Already the purification of the target structure by sublimation at 200°C and a pressure of 2.7·10$^{-2}$ mbar points at its considerably increased vapour pressure in spite of the mass of 1034 amu, a crucial physical property of the compound to obtain sufficient molecular beam intensities for matter wave experiments. However, recrystallization from hot chloroform turned out to be the more efficient purification procedure with an isolated yield of 72%. The new compounds were characterized by NMR spectroscopy and mass spectrometry. Even though the fluorinated azobenzene is a new compound that has been designed and synthesized for this experiment, analogy with comparable derivatives allows to predict particular structural features. Using 'Gaussian 03W V6.0' we determine the scalar molecular polarizability to be 49 A$^3$. Based on earlier comparisons between this program and our own molecule metrology experiments [10] we estimate the error to be smaller than 20%. In addition to their high vapour pressure due to the fluorinated alkyl chains, azobenzene derivatives are known for their reversible photo isomerisation [30] between the thermodynamically favourable *trans*-form (Figure 3b) and the folded *cis*-form.

We used electron impact ionization and quadrupole mass spectrometry (CMS Extrel Merlin 1...4000 u) to detect these molecules after their passage through the interferometer. This allowed to discriminate Δm/m with better than 1% and we still get a sufficient signal for interferometry, under the condition that we open the beam to a vertical height of 400 μm and broaden the velocity spread to Δv/ $v_z$ = 0.25 (standard deviation m/s).

**Author contributions**
The interferometry work was performed by SG, LH, AS, HU, MG, FG, and MA. Analysis was carried out by KH and SG. The nanogratings were prepared by TS in collaboration with LH. SG and LH contributed equally to the experiment. MM and MM designed, synthesized and characterized the fluorinated azobenzene derivate with high vapour pressure.

**Corresponding author**
All correspondence and request of materials should be sent to markus.arndt@univie.ac.at .

**Acknowledgements**
The project is supported by the Austrian FWF within the projects START Y177-2 and SFB F1505, by the European Commission within the RTN network HPRN-CT-2002-00309. KH acknowledges support by the DFG Emmy-Noether program. MM and MM acknowledge support from the Swiss National Science Foundation (SNSF) and the Innovation Promotion Agency (CTI). We thank Anton Zeilinger for the lending of a cw laser.

**Competing interests statement:**
The authors declare that they have no competing financial interests.

**Supplementary Information:** *Synthesis and Characterization of azobenzene 4,4'-di(carboxylic acid 1"H-1"H-perfluoro-1"-octanolate)*

***Reagents and solvents:*** All chemicals were directly used for the synthesis without further purification. The solvents for crystallisation were distilled once before use, the solvents for extraction were used in technical grade. Dry tetrahydrofuran (THF) was distilled over sodium and potassium.

***Synthesis:*** All reactions comprising oxidation or hydrolysis sensitive reagents were done under an argon atmosphere using the Schlenk–technique. Furthermore, dry solvents were used and the glassware was heated out prior to use.

***Analytics and instruments:*** **$^1$H-Nuclear Magnetic Resonance (NMR):** *Bruker DPX-NMR* (400 MHz) and *Bruker BZH-NMR* (250 MHz) instruments were used to record the spectra. Chemical shifts (δ) are reported in parts per million (ppm) relative to residual solvent peaks or trimethylsilyle (TMS), and coupling constants (J) are reported in Hertz (Hz). NMR solvents were obtained from *Cambridge Isotope Laboratories, Inc.* (Andover, MA, USA). The measurements were done at room temperature. The multiplicities are indicated as: s=singlet, d=doublet, q=quartet, quin=quintet, m=multiplet and b=broad.
**$^{13}$C-Nuclear Magnetic Resonance (NMR):** *Bruker DPX-NMR* (400 MHz) instruments were used to record the spectra. Chemical shifts (δ) are reported in parts per million (ppm) relative to residual solvent peaks. NMR solvents were obtained from *Cambridge Isotope Laboratories, Inc.* (Andover, MA, USA). The measurements were done at room temperature. The carbons are classified as: Cp=primary, Cs=secondary, Ct=tertiary, Cq=quaternary.
**Mass spectroscopy (MS):** Mass spectra were recorded on a *Finnigan MAT 95Q* for Electron Impact (EI) or a *Voyager-De™ Pro* for MALDI-TOF. The peaks were measured in m/z (%).
**Melting point (MP):** The melting points were measured on a Stuart melting point apparatus SMP3.

***Azobenzene-4,4'-dicarboxylic acid*** *was synthesized following a reported protocol [1]:*

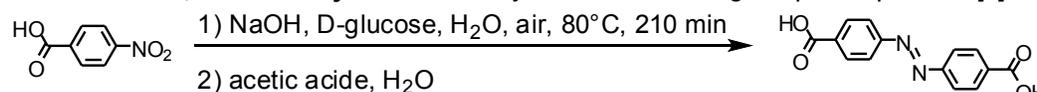

4-Nitrobenzoicacid (1.0 eq., 0.060 mol, 10.0 g) was dissolved in 200 ml water. After adding sodium hydroxide (14.0 eq., 0.838 mol, 33.5 g) portion wise, the solution was heated to 80°C and stirred for 15min. An aqueous solution of D-glucose (5.7 eq., 0.342 mol, 61.5 g) in 120 ml hot water was added drop wise, while the solution turned yellow to brown. Then a strong air stream was passed into the mixture for 210min, after cooling to 0°C a light brown precipitate was obtained. The crude was filtered, dissolved in 400 ml hot water, acidified with 20 ml acetic acid and cooled to 0°C. The light pink precipitate was filtered and dried on high vacuum to give the desired product (16.1g, $C_{14}H_{10}N_2O_4$, 270.24 g/mol, 83% isolated yield).

**$^1$H-NMR** (250 MHz, DMSO-$d_6$, δ/ppm): 13.29 (2H, s, -COO*H*); 8.18 (4H, d, $^3J_{HH}$ = 8.6 Hz, Ar); 8.02 (4H, d, $^3J_{HH}$ = 8.6 Hz, Ar).
**$^{13}$C-NMR** (101 MHz, DMSO-$d_6$, δ/ppm): 167.5 (2C, Cq, s); 155.0 (2C, Cq, s); 134.3 (2C, Cq, s); 131.6 (4C, Ct, s); 123.7 (4C, Ct, s).
**MS** (EI, m/z): 271.0 (8%); 272.0 (15%); 270.0 (48%, M$^+$); 149 (43%); 121 (100%); 65 (33%).

**Azobenzene 4,4'-di(carboxylic acid 1"H-1"H-perfluoro-1"-octanolate)**

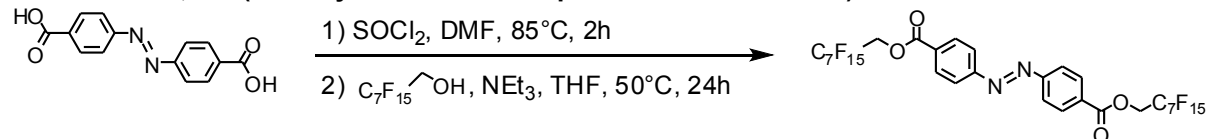

Azobenzene-4,4'-dicarboxylic acid (1.0eq., 5.840mmol, 1.58g) was refluxed in 50ml thionyl chloride containing a few drops of DMF under an argon atmosphere for 2h. The reaction mixture was concentrated to dryness under reduced pressure and dried on high vacuum. The red crude was then dissolved in 80 ml dry THF. This solution was added drop wise to a stirred dry THF solution (80 ml) of 1H-1H-perfluoro-1-octanol (2.1 eq., 12.500 mmol, 5.0 g) and triethylamine (5.0 eq., 29.200 mmol, 2.95 g) at 0°C under an argon atmosphere. After stirring for 24h at 50°C the reaction mixture was evaporated to dryness and the red residue dissolved in 30ml DMF and poured into 300ml ice water. After 30min at -20°C the precipitate was filtered and dried on high vacuum.

The light red crystals was sublimated at 200°C at 2.7·10$^{-2}$ mbar to give the product as orange crystals (1.4 g, $C_{30}H_{12}F_{30}N_2O_4$, 1034 g/mol, 23%). In a second attempt the light red crystals was recrystallized from 400ml chloroform to give the product as orange crystals (4.3g, $C_{30}H_{12}F_{30}N_2O_4$, 1034g/mol, 72%).

**$^1$H-NMR** (400 MHz, CDCl$_3$, δ/ppm): 8.24 (4H, d, $^3J_{HH}$ = 8.5 Hz, Ar); 8.03 (4H, d, $^3J_{HH}$ = 8.5 Hz, Ar); 4.88 (4H, t, $^3J_{HH}$ = 13.3 Hz, -C*H$_2$*-CF$_2$-).

**$^{19}$F-NMR** (400 MHz, CDCl$_3$, δ/ppm): -81.8 (3F, t); -120.3 (2F, m); -123.1 (4F, m); -123.8 (2F, m); -124.3 (2F,m); -127.2 (2F, m).

**MS** (MALDI-TOF, m/z): 1034 (100%); 1036 (100%).

**MP** 154-156°C